\documentclass[journal,10pt]{IEEEtran}
\IEEEoverridecommandlockouts

\usepackage{array}
\usepackage{amsbsy}
\usepackage{amsthm}
\usepackage{amssymb}
\usepackage{amsmath}
\usepackage{multirow}
\usepackage{epsfig}
\usepackage{subfigure}
\usepackage{graphics}
\usepackage{multicol}
\usepackage{diagbox}
\usepackage{upgreek}
\usepackage{threeparttable}

\usepackage{flushend}

\usepackage{epstopdf}

\usepackage{makecell}
\usepackage[]{pifont}
\usepackage{epstopdf}
\usepackage{balance}
\usepackage{xcolor}
\usepackage{cite}
\usepackage{color}
\usepackage{algorithm}
\usepackage{algorithmic}
\usepackage{textcomp}
\usepackage{subfigure}
\usepackage{subfloat}
\usepackage{booktabs}
\usepackage{lipsum}
\usepackage{bm}
\usepackage{filecontents}

\usepackage{graphicx}

\makeatletter
\newcommand*\bigcdot{\mathpalette\bigcdot@{.5}}
\newcommand*\bigcdot@[2]{\mathbin{\vcenter{\hbox{\scalebox{#2}{$\m@th#1\bullet$}}}}}
\makeatother

\usepackage{scalerel}
\usepackage{stackengine}
\usepackage{pgf}
\newcounter{iloop}
\newcommand\openbigstar[1][0.7]{%
  \scalerel*{%
    \stackinset{c}{-.125pt}{c}{}{\scalebox{#1}{\color{white}{$\bigstar$}}}{%
      $\bigstar$}%
  }{\bigstar}
}
\newcommand{\Stars}[1]{\ensuremath{%
\pgfmathtruncatemacro{\imax}{ifthenelse(int(#1)==#1,#1-1,#1)}%
\pgfmathsetmacro{\xrest}{0.9*(1-#1+\imax)}%
\setcounter{iloop}{0}%
\loop\stepcounter{iloop}\ifnum\value{iloop}<\the\numexpr\imax+1
\bigstar\repeat
\openbigstar[\xrest]%
\setcounter{iloop}{0}%
\loop\stepcounter{iloop}\ifnum\value{iloop}<\the\numexpr5-\imax\relax
\openbigstar[.9]\repeat}}

\usepackage{colortbl}
\usepackage{xcolor}

\graphicspath{ {figures/}}

\usepackage{hyperref}
\hypersetup{
    colorlinks=true,
    linkcolor=blue,
    filecolor=blue,      
    urlcolor=blue,
    citecolor=cyan,
}

\pdfoutput=1
\begin{document}

\title{{\huge Sensing With Random Communication Signals}}

\author{Shihang Lu,~\IEEEmembership{Graduate Student Member,~IEEE},  Fan Liu,~\IEEEmembership{Senior Member,~IEEE}, Yifeng Xiong,~\IEEEmembership{Member,~IEEE}, \\ Zhen Du,~\IEEEmembership{Member,~IEEE}, Yuanhao Cui, \IEEEmembership{Member,~IEEE}, Shuangyang Li, \IEEEmembership{Member,~IEEE}, \\ Weijie Yuan,~\IEEEmembership{Senior Member,~IEEE}, Jie Yang,~\IEEEmembership{Member,~IEEE},
and Shi Jin,~\IEEEmembership{Fellow,~IEEE}

\vspace{-1.5em}

\thanks{S. Lu, F. Liu, and S. Jin are with the National Mobile Communications Research Laboratory, Southeast University, Nanjing 210096, China. 
Y. Xiong and Y. Cui are with the School of Information and Communication Engineering, Beijing University of Posts and Telecommunications, Beijing 100876, China.
Z. Du is with the School of Electronic and Information Engineering, Nanjing University of Information Science and Technology, Nanjing 210044, China.
S. Li is with the Faculty of Electrical Engineering and Computer Science, Technical University of Berlin, Berlin, 10587, Germany.
W. Yuan is with the School of System Design and Intelligent Manufacturing (SDIM), Southern University of Science and Technology, Shenzhen 518055, China.
J. Yang is with the Frontiers Science Center for Mobile Information Communication and Security and also with the Key Laboratory of Measurement and Control of Complex Systems of Engineering, Southeast University, Nanjing 211189, China.
(\textit{Corresponding author: Fan Liu})}
}

\maketitle

\begin{abstract}
Communication-centric Integrated Sensing and Communication (ISAC) has been recognized as a promising methodology to implement wireless sensing functionality over existing network architectures, due to its cost-effectiveness and backward compatibility to legacy cellular systems. However, the inherent randomness of the communication signal may incur huge fluctuations in sensing capabilities, leading to unfavorable detection and estimation performance. To address this issue, we elaborate on random ISAC signal processing methods in this article, aiming at improving the sensing performance without unduly deteriorating the communication functionality. Specifically, we commence by discussing the fundamentals of sensing with random communication signals, including the performance metrics and optimal ranging waveforms. Building on these concepts, we then present a general framework for random ISAC signal transmission, followed by an in-depth exploration of time-domain pulse shaping, frequency-domain constellation shaping, and spatial-domain precoding methods. We provide a comprehensive overview of each of these topics, including models, results, and design guidelines. Finally, we conclude this article by identifying several promising research directions for random ISAC signal transmission.
\end{abstract}

\IEEEpeerreviewmaketitle

\section{Introduction}\label{Introduction}

\subsection{Backgrounds: ISAC Signal Design Methods}

With the advanced evolution of wireless systems, future communication networks are expected to integrate radar sensing functionalities and further achieve Integrated Sensing and Communication (ISAC). It is well-recognized that ISAC may play a critical role in enabling various emerging applications such as vehicle-to-everything (V2X) and low-altitude intelligent networks \cite{keskin2024SPM,marwa_tuts}. Moreover, the International Telecommunication Union (ITU) has listed ISAC as one of the six typical usage scenarios for future sixth-generation (6G) networks. Meanwhile, the European Telecommunications Standards Institute (ETSI) has established an Industry Specification Group for ISAC, aiming to lay the technical foundation for standardizing ISAC technologies. These practical developments highlight the growing necessity of integrating sensing capabilities into existing cellular networks \cite{Zhiqing_netw, mishra2019SPM}.

The concept of ISAC originates from radar and communication spectrum sharing (RCSS), which seeks to utilize the same frequency bands for both radar sensing and communication (S\&C) tasks. Therefore, ISAC may alleviate spectrum congestion and enhance overall spectrum efficiency. Beyond mere spectrum sharing, the ultimate goal of ISAC systems is to emit unified waveforms that achieve simultaneous and seamless S\&C functionalities, thereby maximizing the resource efficiency \cite{OJCOMS_2023,duzhenTSP,zhangyumeng2023input}. Generally, there are three primary design methodologies for ISAC signaling: sensing-centric, communication-centric, and joint signal design. As illustrated in Table \ref{tab1}, we elaborate on both the advantages and disadvantages of the above approaches:

\subsubsection{Sensing-Centric ISAC Signaling} Sensing-centric design, also termed as radar-centric ISAC, primarily relies on deterministic signals, such as the commonly used frequency-modulated continuous-wave (FMCW) and phase-modulated continuous-wave (PMCW) \cite{mishra2019SPM}. These pure deterministic signals enable communication information to be embedded within legacy radar pulses, e.g., chirp signals, which typically employ slow-time coding by utilizing inter-pulse modulation techniques. Consequently, the communication rate is limited by the radar's pulse repetition frequency (PRF), resulting in a reduced transmission rate, even though the sensing performance remains reliable. More severely, this approach is not compatible with 3GPP standards, and is thus challenging to implement in existing cellular networks. 

\subsubsection{Communication-Centric ISAC Signaling} 
In contrast to its sensing-centric counterpart, the communication-centric design realizes the radar sensing functionality by leveraging standard communication waveforms, such as orthogonal frequency division multiplexing (OFDM) signals \cite{zhangyumeng2023input,keskin2024fundamental,duzhenTSP}. The key advantage lies in maintaining the communication quality of cellular users while fully reusing spectrum resources, providing sensing as a basic service to users \cite{Zhiqing_netw}. However, it is worth noting that communication signals have to be random to effectively convey useful information, which may introduce fluctuations in radar sensing performance. As a result, classical sensing performance metrics defined upon the pure deterministic signals, e.g., Cr\'amer-Rao bound (CRB) and ambiguity function (AF), may exhibit random variations across different signal realizations, leading to unreliable sensing performance.

\begin{table*}[!ht]
\centering
\begin{tabular}{>{\columncolor{teal!20}}m{3.3cm} >{\columncolor{gray!20}}m{4.2cm} >{\columncolor{teal!20}}m{4.3cm} >
{\columncolor{gray!20}}m{2.6cm} >{\columncolor{teal!20}}m{1.7cm} }
Design Methodology & Key Features & Pros \& Cons & Examples & Compatibility \\
\hline
Sensing-Centric Design  & \makecell[l]{$\bigcdot$ Deterministic Signals\\ $\bigcdot$ Information Embedded    Pulses \\ $\bigcdot$ Inter-Pulse Modulation }   & \makecell[l]{$\bigcdot$ Reliable Sensing Performance \\ $\bigcdot$  Low Communication Rate \\ $\bigcdot$ Low Spectrum Efficiency} & \makecell[l]{$\bigcdot$ FMCW \cite{mishra2019SPM} \\ $\bigcdot$ PMCW \cite{mishra2019SPM}} & \Stars{3}   \\
\specialrule{0em}{.2pt}{.2pt}
Commun.-Centric Design& \makecell[l]{$\bigcdot$ Random Signals \\$\bigcdot$ Pilot Signals\\ $\bigcdot$ Reusing Data Payloads \\ $\bigcdot$ Commun.-Specific Waveform} & \makecell[l]{$\bigcdot$  Low Cost \\ $\bigcdot$ High Spectrum Efficiency \\ $\bigcdot$ Degraded Sensing Performance}  & \makecell[l]{$\bigcdot$ OFDM \cite{Yuguangding_OFDM,zhangyumeng2023input,keskin2024fundamental,duzhenTSP} \\  $\bigcdot$ OTFS \cite{OTFS2024}} &  \Stars{5} \\
\specialrule{0em}{.2pt}{.2pt}
Joint Design & \makecell[l]{$\bigcdot$ Novel Signaling Formats\\ $\bigcdot$ Iterative Algorithms \\ $\bigcdot$ Pareto Optimization}  & \makecell[l]{$\bigcdot$ Scalable S\&C Tradeoff \\$\bigcdot$ Bound-Achieving Strategies \\ $\bigcdot$ High Complexity}  & \makecell[l]{$\bigcdot$ CRB-SINR \\~~Tradeoff Design \cite{Zhiqing_netw} \\ $\bigcdot$ Beampattern \\ ~~Design Method \cite{Zhiqing_netw}} & \Stars{1}\\
\specialrule{0em}{.2pt}{.2pt}
\end{tabular}
\begin{small}\caption{The comparasions among existing ISAC signal design methods.} \label{tab1} \end{small}
\end{table*}

\subsubsection{Joint-Design ISAC Signals} Both the sensing- and communication-centric designs prioritize only one type of S\&C performance while failing to achieve a scalable tradeoff between the two. In contrast, the joint-design method enables a favorable balance between S\&C performance by formulating tractable optimization problems. For example, recent state-of-the-art works aim to minimize a weighted loss function containing both S\&C performance metrics under practical constraints. By doing so, one may attain the Pareto-optimal boundary by tuning the weight parameters \cite{Zhiqing_netw}. Nevertheless, such tailored waveforms are typically generated through computationally expensive iterative algorithms, leading to considerable implementation challenges in future 6G networks.

\subsection{Motivations: Utilizing Random Communication Signals}

Considering the compatibility with existing standards, the communication-centric ISAC design emerges as the most promising approach for integrating sensing capabilities into current cellular networks by fully reusing the transmission hardware and wireless resources, as compared with sensing-centric and joint design methods. Nevertheless, the current ISAC signaling methods based on the fifth-generation new radio (5G NR) utilize only a small portion of physical layer reference signals, such as channel statement information reference signals (CSI-RS). These deterministic reference signals, while possessing favorable auto-correlation properties, occupy merely 10\% of the signal frame structure, which is far away from sufficient for implementing the high-accuracy sensing functionality required in future 6G networks. To that end, one has to repurpose the remaining 90\% of time-frequency resource blocks, embedded with random data payloads, to enhance sensing capabilities and improve the resource efficiency.

In sharp contrast to the deterministic reference signals, data payloads are supposed to be randomly drawn from predefined codebooks to carry useful information, e.g., Gaussian codebooks and discrete Quadrature Amplitude Modulation (QAM)/Phase-Shift Keying (PSK) alphabets. As elaborated above, the communication rate in such a case is guaranteed, at the cost of degraded sensing performance due to the randomness of communication signals. In this context, one needs to conceive tailored signal processing methods to cope with the inherent randomness of data payload signals, thus improving the sensing performance, which motivates this article. 

In the following sections, we commence with the sensing metrics under random signaling and then discuss the optimal communication-centric ranging waveform in Sec. \ref{Sec2}. Next, we outline a general random ISAC signal processing framework and provide case studies to demonstrate the performance optimization in Sec. \ref{Sec3}. Within this unified framework, we explore time-domain pulse shaping, frequency-domain constellation shaping, and spatial-domain multiple-input multiple-output (MIMO) precoding, respectively. The related results validate the effectiveness of the proposed random ISAC signal processing methods while offering valuable design insights. Finally, we summarize future directions on random ISAC signal processing deserving further investigation in Sec. \ref{Sec4}.

\section{Sensing with Random Communication Signals}\label{Sec2}

In this section, we attempt to answer two fundamental questions in sensing with random signals: \textit{How to characterize the sensing performance under random signaling?}
\textit{What is the optimal communication-centric ranging waveform?} This offers valuable design guidance for ISAC system implementation, as detailed in the following.

\subsection{How to Characterize the Sensing Performance Under Random Signaling?}

Classical radar performance metrics (e.g., detection probability, mean squared error (MSE), CRB, and AF) are mainly defined upon deterministic signaling and may not be suitable when leveraging random ISAC signals. This is because the random data payload causes variability of sensing performance metrics during each signal transmission. Therefore, the primary challenge is to accurately characterize the sensing performance considering the signal randomness.

For clarity, let us denote the deterministic radar signals by $\bm{S}_D$ and the to-be-optimized parameter (e.g., precoding matrix) by $\bm{W}$. Traditional radar performance metrics are typically represented by specific loss functions, denoted as $f(\bm{W}; \bm{S}_D)$, such as CRB, linear squared error (LSE), and linear minimum mean square error (LMMSE), detection probability, and false-alarm probability \cite{LuTSP2024,biguesh2006training, AFDM_twc}. Notice that $f(\bm{W}; \bm{S}_D)$ is a deterministic function of both $\bm{W}$ and $\bm{S}_D$. In the MIMO radar case where $\bm{S}_D$ is an orthogonal signaling matrix, the sensing loss function $f(\bm{W}; \bm{S}_D)$ depends only on the precoding matrix $\bm{W}$ \cite{LuTSP2024,biguesh2006training}.

\begin{figure*}[!ht]
\centering
\subfigure[] { \label{Phy1}
			\includegraphics[width=0.205\textwidth]{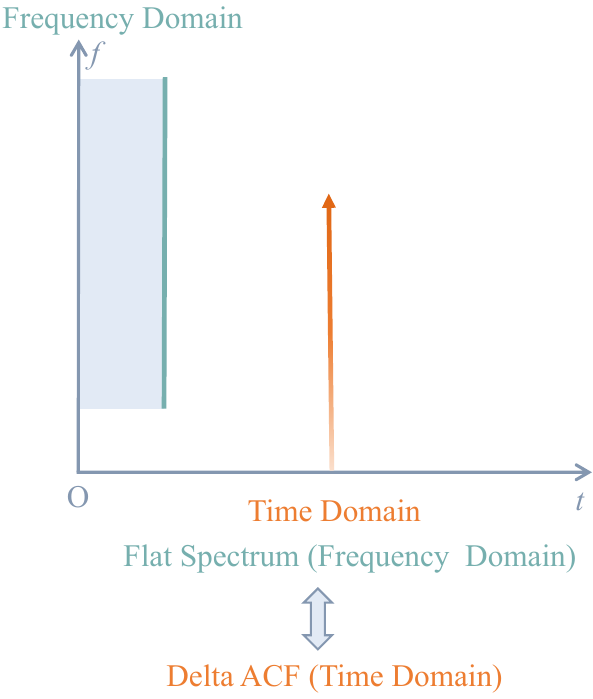}
		}
\subfigure[] { \label{Phy2}
			\includegraphics[width=0.205\textwidth]{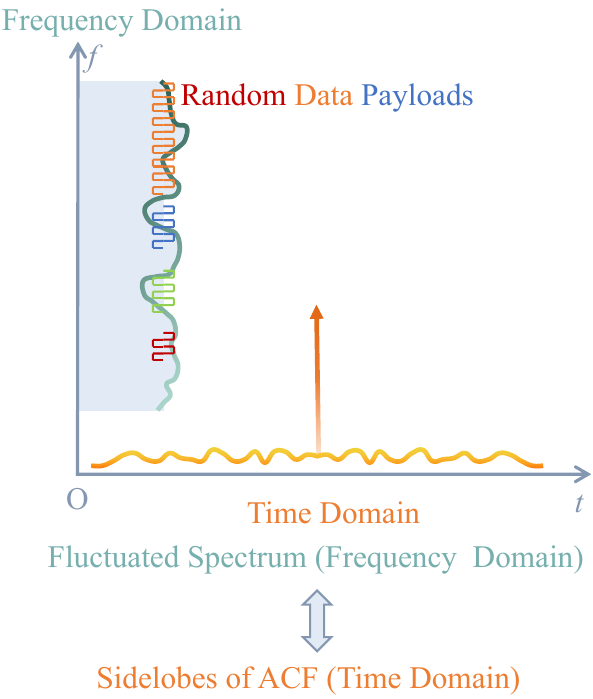}
		}
\subfigure[] { \label{Phy3}
			\includegraphics[width=0.205\textwidth]{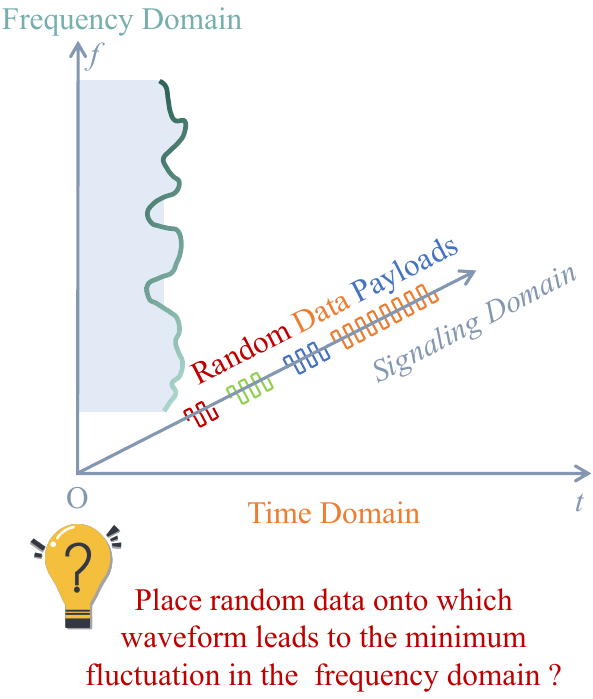}
		}
\subfigure[] { \label{sidelobe_compare}
			\includegraphics[width=0.30\textwidth]{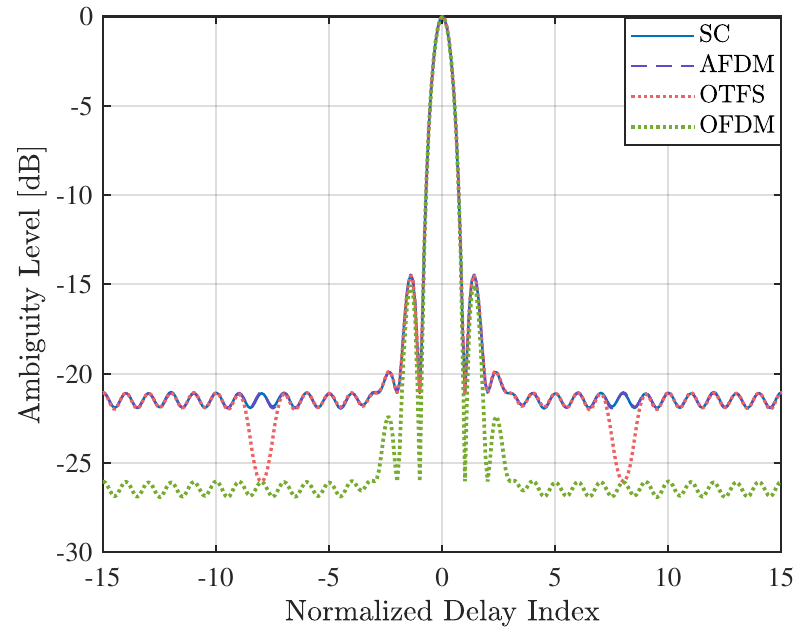}
		}
\caption{Physical interpretation about the optimal ranging performance of OFDM and SC modulation.(a) The ideal case: A Dirac-Delta ACF; (b) Practical case: Randomness causes variability; (c) Waveform Design Insights: Where to place random data? (d) The sidelobe comparisons among SC, AFDM, OTFS, and OFDM waveforms, under 16-QAM constellation and root-raised cosine (RRC) pulse-shaping filter with roll-off 0.35.} 
        \label{Phy}
\end{figure*}

Things come differently when random ISAC signals are employed. In this case, the sensing performance metrics are primarily influenced by each specific realization of the transmitted random signal. Accordingly, the instantaneous sensing loss function is re-expressed as $f(\bm{W}; \bm{S}), \bm{S} \in \mathcal{A}$, where $\bm{S}$ represents a random signal matrix, whose entries are drawn from a given alphabet $\mathcal{A}$ (e.g., Gaussian codebooks or QAM/PSK symbol sets). Accordingly, $f(\bm{W}; \bm{S})$ is treated as a random function with respect to $\bm{S}$.
In this context, $f(\bm{W}; \bm{S})$ may not be an adequate metric to quantify the sensing performance, as it depends only on a single realization of $\bm{S}$. To tackle this issue, there are two main approaches to quantifying sensing performance under random signaling: one focuses on instantaneous outage performance, while the other emphasizes average sensing performance, as detailed below:

\textbf{Outage Performance:} Due to the randomness of the signal $\bm{S}$, $f(\bm{W}; \bm{S})$ follows a certain probability distribution. The sensing outage probability may be defined as the tail probability of $f(\bm{W}; \bm{S})$, i.e., the probability that $f(\bm{W}; \bm{S})$ exceeds (or falls below) a certain threshold $f_{0}$, denoted by $\mathbb{P}[f(\bm{W}; \bm{S}) \ge f_{0}]$. For example, in an OFDM-ISAC system, this may correspond to the probability of incorrectly estimating the delay-Doppler bin for any given $\bm{S}$ realization \cite{zhangyumeng2023input}. To evaluate the sensing reliability, $f_0$ may represent the MSE threshold, where the outage events occur if the instantaneous MSE is greater than $f_0$. Another example is found in detection tasks, where $f_0$ may represent the SNR threshold, sensing rate requirement, or desirable detection probability threshold \cite{AFDM_twc}.

\textbf{Average Performance:} Note that most sensing scenarios involve continuous, long-term detection and estimation tasks, e.g., multiple and continuous range estimations of vehicles in V2X scenarios. Then, the focus should be on the average performance over a prolonged period rather than the specific realization of a single signal instance $\bm{S}$. In this context, it is straightforward to take the expectation of $f(\bm{W}; \bm{S})$ over $\bm{S}$ that is $\mathbb{E}_{\bm{S}}[f(\bm{W}; \bm{S})]$, thereby enabling further analysis of the average sensing performance. For example, $\mathbb{E}_{\bm{S}}[f(\bm{W}; \bm{S})]$ may represent the average ambiguity function (AF) \cite{duzhenTSP}, the expectation of the integrated sidelobe level (EISL) \cite{liu2025uncovering}, sensing mutual information (SMI), and so on \cite{LuTSP2024}.

\begin{figure*}[ht!]
    \centering
    \includegraphics[width=0.95\linewidth]{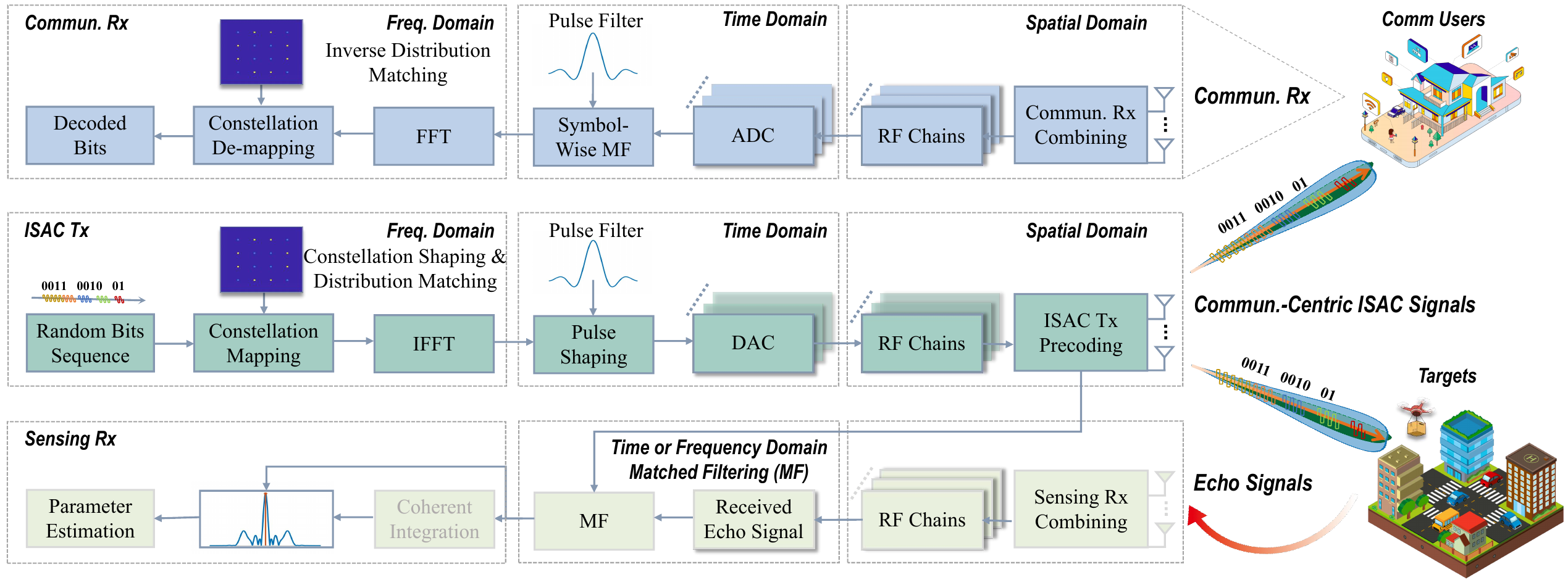}
    \caption{A general framework of communication-centric ISAC transmission under OFDM signaling. ADC: analog-to-digital converter; DAC:  digital-to-analog converter.}
    \label{FrameWork}
\end{figure*}

\subsection{What is the Optimal Communication-Centric Ranging Waveform?}

Communication-centric ISAC waveforms mainly leverage standardized waveforms in current wireless systems, such as single-carrier (SC), OFDM, and code-division multiple access (CDMA) waveforms \cite{liu2024ofdm}. Additionally, emerging waveform paradigms like orthogonal time-frequency space (OTFS) and affine frequency-division multiplexing (AFDM) are being explored as promising candidates to address high-mobility issues in future networks \cite{OTFS2024, AFDM_twc}. All the aforementioned waveforms employ a well-designed orthonormal basis to modulate random data symbols, with the waveform type represented by an unitary modulation basis matrix (also called transformation matrix). Accordingly, the study of optimal communication-centric ISAC waveforms essentially boils down to investigating the signaling basis matrix. Below we elaborate on the structure of modulation bases for commonly used waveforms.

\textbf{SC:} The modulation basis matrix of SC signals is simply an identity matrix that conveys data symbols directly in the time domain. This matrix consists solely of discrete impulse functions, indicating that the SC signal samples are essentially a series of weighted discrete Kronecker-Delta functions in the time domain.

\textbf{OFDM:} The random data symbols are placed in the frequency-time domain. The waveform basis consists of discrete complex exponential functions, forming an inverse discrete Fourier transform (IDFT) matrix that provides orthogonality among subcarriers. After performing the IDFT on the frequency-domain random symbols, OFDM signals are represented as a ``weighted superposition'' of multiple sinusoidal components in the time domain, where the ``weights'' are exactly the random data symbols. 

\textbf{OTFS:} The modulation basis consists of two-dimensional localized functions in the delay-Doppler domain, represented by the inverse discrete Zak transform (IDZT) matrix. As a two-dimensional modulation scheme, OTFS maps data symbols onto a delay-Doppler grid, leveraging the time-varying characteristics of wireless channels.

\textbf{AFDM:} The transformation matrix represents the inverse discrete affine Fourier transform (IDAFT), which is used to map symbols from the affine Fourier transform (AFT) domain to the time domain for wireless transmission. Additionally, AFDM facilitates the reconstruction of the delay-Doppler representation of wireless channels, with fine-tuned chirp parameters adapting to the channel characteristics.

\textbf{Optimal Ranging Waveform:} For mono-static ISAC transmission operated in static/quasi-static environments, it is rigorously revealed that for all sub-Gaussian constellations (e.g., QAM and PSK), the OFDM signal is the unique optimal waveform that achieves the average lowest ranging sidelobe \cite{liu2024ofdm}. As illustrated in Fig. \ref{Phy}, we provide a visual physical interpretation to explain the essence of the optimal ranging waveform. We have the following observations:

\begin{itemize}
    \item \textbf{Ideal Case:} In the matched filtering scheme, an optimal ranging waveform seeks to achieve an ideal Dirac-Delta function as the output of the matched filter, thereby eliminating any multi-target ambiguity. As depicted in Fig. \ref{Phy1}, a perfectly uniform frequency spectrum with infinite bandwidth corresponds to a Dirac-Delta function in the time domain. This outcome is highly desirable, as the ideal Dirac-Delta function ensures a perfectly matched filter response.
    
    \item \textbf{Practical Case:} As shown in Fig. \ref{Phy2}, due to the modulated random data payloads, both the time-domain and frequency-domain signals may experience undesirable fluctuations, which can lead to imperfect MF output. Consequently, the random data signals introduce random sidelobes in the autocorrelation function (ACF), jeopardizing the ranging performance, particularly in multi-target scenarios. According to the Wiener-Khinchin theorem, the Fourier transform of the time-domain ACF is equal to the power spectral density (PSD) of the frequency-domain signal. Therefore, minimizing the average sidelobes of the time-domain ACF may be achieved by reducing the power fluctuations of the signal, which corresponds to minimizing its fourth-order moment (a.k.a. kurtosis) in the frequency domain \cite{liu2024ofdm}.

    \item \textbf{Design Insights:} The modulation basis may be treated as an unitary transform over i.i.d. random data symbols. According to the central limit theorem (CLT), linear transformations of a random vector with i.i.d. entries may cause the vector to asymptotically approach a Gaussian distribution. For sub-Gaussian constellations of the frequency domain, multiplying the waveform basis matrix, increase the kurtosis. Therefore, it attains optimal ranging performance by retaining its kurtosis of the frequency domain. This requires that the product of the DFT matrix and the modulation basis matrix be the identity matrix, which corresponds to the IDFT operation. Therefore, this leads to the OFDM waveform.
\end{itemize}

\begin{figure*}[ht!]
\centering
\subfigure[Time-domain pulse shaping design.] { 
\label{PulseShaping_Scenario}
			\includegraphics[width=0.40\textwidth]{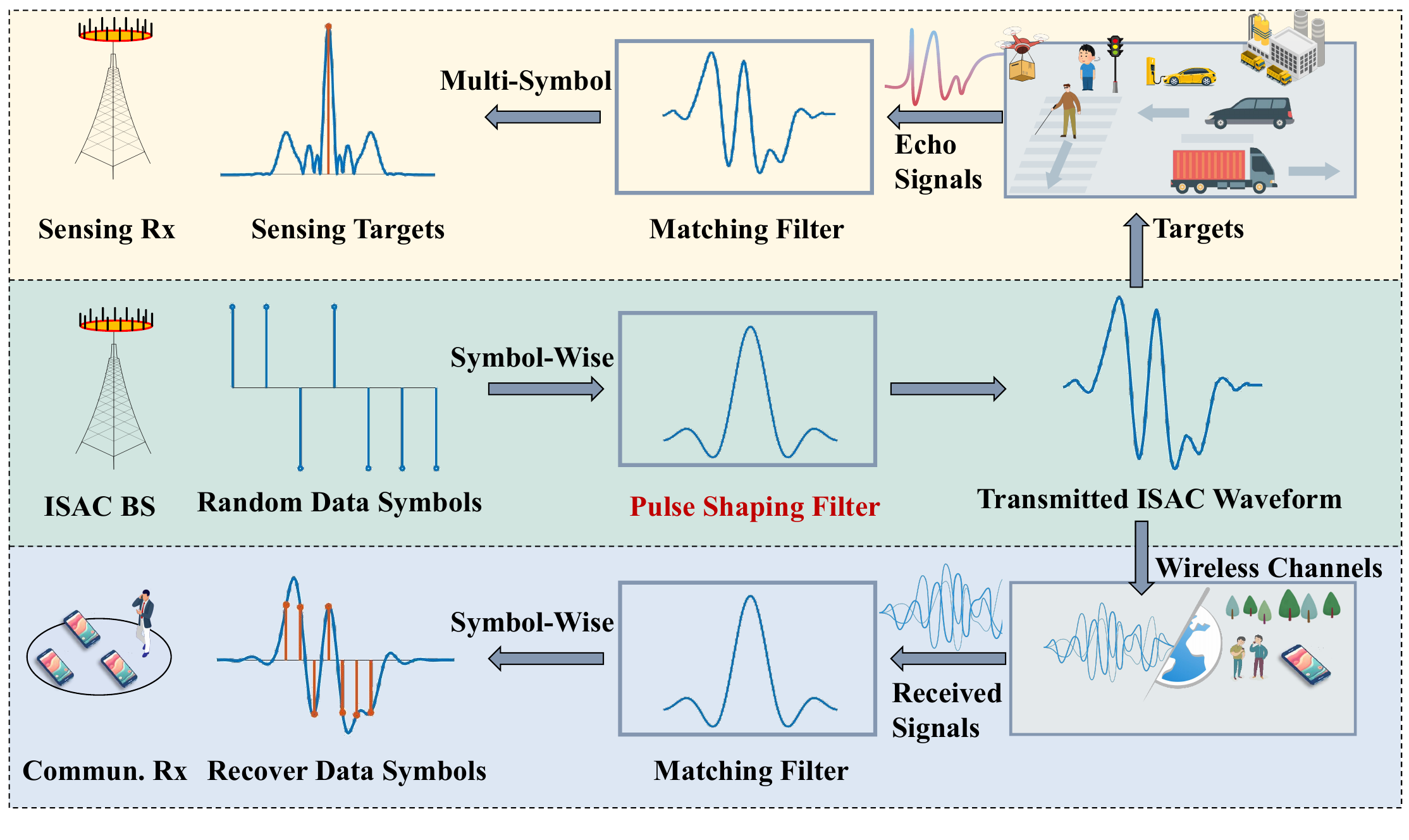}
		}
\subfigure[Sensing performance comparisons.] { 
\label{PS_Performance}
            \includegraphics[width=0.272\linewidth]{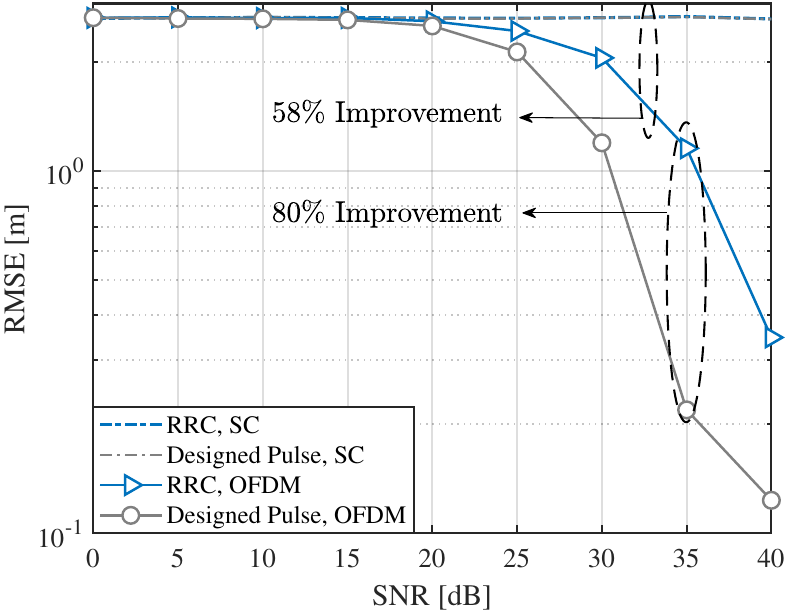} 
            }
\subfigure[S\&C performance tradeoffs.] { 
\label{PS_Rangeprofile}
            \includegraphics[width=0.272\textwidth]{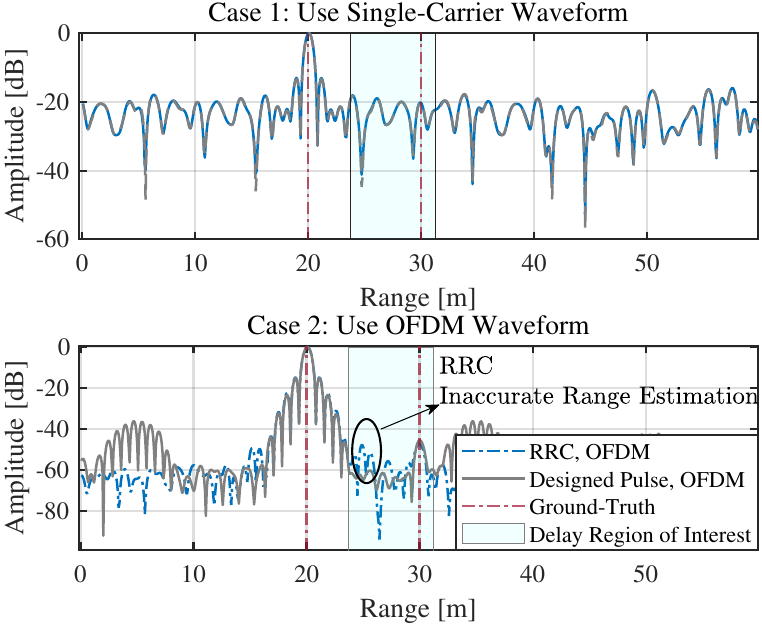} }
\caption{Time-domain pulse shaping design in OFDM-ISAC systems: Beneficial results in sensing.}
        \label{PulseShaping}
\end{figure*}

\section{Random ISAC Signal Processing}\label{Sec3}

In this section, we first establish a general framework for random ISAC signal processing, clarifying the key parameters and techniques that may be optimized and processed. Subsequently, we present the methods and beneficial improvements of random ISAC signal processing from the perspectives of time, frequency, and spatial domains, respectively.

\subsection{General Framework}

Given the optimality of OFDM waveform in ranging as well as its compatibility with existing networks, let us examine an OFDM-ISAC system in this section. As shown in Fig. \ref{FrameWork}, we present a general mono-static ISAC framework toward random ISAC signal processing, comprising an ISAC transmitter (Tx) and S\&C receiver (Rx). The key processing units of the ISAC transmitter are divided into three parts: frequency, time, and spatial domain, which are interconnected to support both communication and sensing functionalities. It is worth pointing out that ISAC transmission is entirely built upon communication systems, with no additional dedicated sensing units required. The only cost incurred may be the addition of a dedicated ISAC Rx. As a result, this significantly reduces the hardware cost of implementing ISAC and greatly enhances resource efficiency. Note that the signal flow begins with the generation of random data for communication, followed by several signal processing stages, including constellation shaping, pulse shaping, and radio frequency (RF) chains. The communication-centric ISAC signal is then emitting at the ISAC Tx and subsequently received at the S\&C receivers, respectively. Due to the randomness of ISAC signals, it is critical to optimize the above transmission blocks to attain favorable sensing performance, as examined below.

\subsection{Time Domain: Pulse Shaping}

\subsubsection{Backgrounds} In classical OFDM systems, each symbol is modulated using a rectangular pulse in the time domain. The rectangular pulse corresponds to a sinc-shaped spectrum in the frequency domain, which exhibits slow decay and results in high out-of-band emissions (OOBE). To that end, pulse shaping method may be leveraged to restrict the spectrum leakage of the signal, while eliminating the inter-symbol interference (ISI). This process essentially involves convolving each time-domain OFDM symbol by the well-designed filter, where each random communication symbol is assigned the shape of the pulse. At the communication receiver, matched filtering is performed by again convolving the received signal with the pulse, followed by the detection of the data symbols through sampling at the integer multiple of the symbol duration. Therefore, pulse shaping is a symbol-wise operation for communications. However, the sensing Rx needs to perform a correlation between the received echo signals and the entire transmitted ISAC signals (rather than just the pulse) to achieve matched filtering and target parameter estimation. Given the above backgrounds, notice that the pulse shape may affect both sensing and communication performance. To boost sensing performance without degrading the communication performance, it is essential to design a Nyquist pulse with favorable ambiguity properties tailored for sensing.  

\subsubsection{Models and Solutions} 
As illustrated in Fig. \ref{PulseShaping}, the ISAC BS is assumed to utilize random signaling to estimate the range of multiple targets while simultaneously communicating with users. To achieve accurate range estimation for weak targets, it is essential to prevent the sidelobes of strong targets from overshadowing the mainlobes of the weak targets. 
A closer examination on the expectation of the ACF under i.i.d. constellations and Nyquist pulse shaping filters indicate that the sensing performance hinges on both the kurtosis of the constellation and the pulse shape, which determine the variance and the mean of the ACF, respectively. In the case that a coherent integration operation is performed, which reduces the variance of the ACF, the ranging error is attributed mainly by the sidelobe level of the ACF of the pulse itself. This may be minimized through convex optimization approaches subject to the Nyquist conditions \cite{liu2025uncovering}.

\subsubsection{Main Results} In Fig. \ref{PS_Performance}, we demonstrate the effectiveness of the proposed pulse shaping design methods in ranging performance. We compare the designed pulse with RRC pulse, under SC and OFDM waveforms, respectively. Assume there is a strong target at 20m and a weak target at 30m. The designed Nyquist pulse effectively ranges the weak target by minimizing the ISLR within the interested delay region of [23.74m, 31.24m]. It is observed that OFDM significantly outperforms SC waveform in ranging performance, which is aligned with our analysis in Sec. \ref{Sec2}. For the RRC pulse, there is an approximately 58\% performance improvement under OFDM waveforms. Furthermore, the designed pulse significantly outperforms the RRC pulse. Under OFDM waveforms in Fig . \ref{PS_Rangeprofile}, it is shown that the designed pulse attains lower sidelobe levels as compared to the RRC pulse, suggesting that tailored pulse shaping is superior in detecting weak targets in the interested delay region. The related results indicate that the proposed pulse shaping methods may attain favorable ranging performance under random signaling.

\subsubsection{Design Guidelines} From the perspective of compatibility with existing networks, the proposed pulse shaping scheme is compatible with existing filtered-OFDM (F-OFDM) systems. At the communication receiver, the pulse selection may be configured by the protocols to achieve MF and recover the random data. Under random ISAC signaling, the pulse shaping methods may be sophisticatedly conceived to attain reliable sensing performance without compromising the communication rate.

\subsection{Frequency Domain: Constellation Shaping}

\subsubsection{Backgrounds} The optimal constellations of S\&C are much different from each other. Considering a single-input and single-output (SISO) Gaussian channel, it is revealed that a constant-amplitude constellation (e.g., PSK) is sensing optimal, whereas the constellation ideal for communication is a Gaussian distribution. As a consequence, the S\&C performance may achieve a beneficial tradeoff by carefully designing the input distribution of the constellation. There are generally three key shaping techniques: probabilistic constellation shaping (PCS), geometric constellation shaping (GCS), and joint constellation shaping (JCS). The PCS method keeps the positions of the constellation points unchanged, while modifying the transmission probability of constellation points to create a non-uniform distribution. Alternatively, GCS involves optimizing the positions of constellation points while keeping their probabilities unchanged. As a combination of PCS and GCS, the JCS method aims to jointly optimize both probabilities and locations of the constellations. 

\subsubsection{Models and Solutions} As a basic example to illustrate the effectiveness of the PCS method, let us consider a monostatic ISAC system, where the ISAC-BS emits OFDM signals for achieving both S\&C tasks \cite{duzhenTSP}. For sensing tasks, notice that the OFDM symbols may cause random sidelobes of the AF, leading to erroneous target detection and parameter estimation. As mentioned above, the AF variance depends on the kurtosis of the constellation types. Therefore, the PCS method may be implemented by optimizing the probability of each constellation point, while maximizing the communication rate subject to kurtosis and transmit power constraints. The formulated optimization problem may be optimally solved by a modified Blahut-Arimoto algorithm \cite{duzhenTSP}. 

\begin{figure}[t!]
\centering
\subfigure[64-QAM-PCS results under different kurtosis constraints.] { \label{64QAM_PCS}
			\includegraphics[width=1\linewidth]{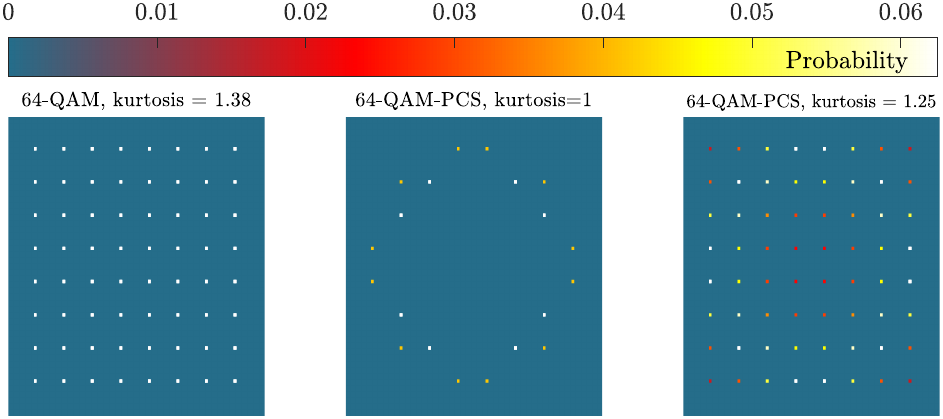}
		}
\subfigure[S\&C tradeoffs using 64-QAM-PCS.] { \label{64QAM_Tradeoffs}
\includegraphics[width=1\linewidth]{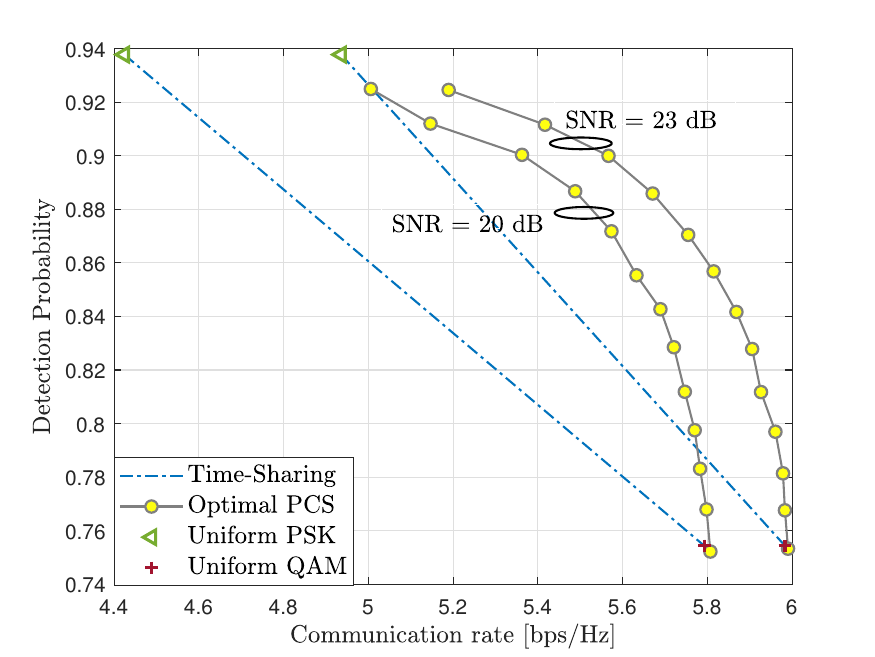} }
\caption{The PCS results and ISAC tradeoffs using 64-QAM constellation.}
        \label{16QAM_PCS_all}
\end{figure}

\begin{figure*}[h!]
\subfigure[The considered MIMO-ISAC systems.] { 
\label{Precoding_Scenario}
			\includegraphics[width=0.40\textwidth]{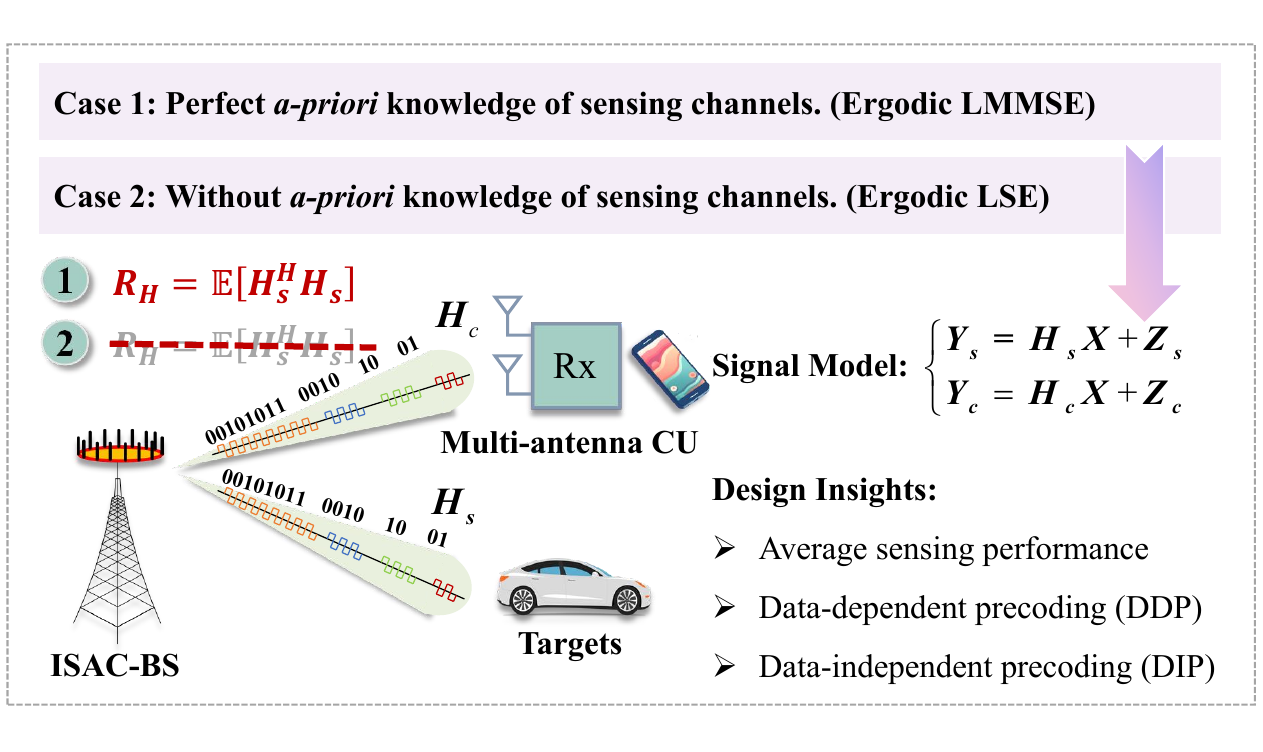}
		}
\subfigure[Sensing performance comparisons.] { 
\label{ELSE_SensingOnly}
            \includegraphics[width=0.272\linewidth]{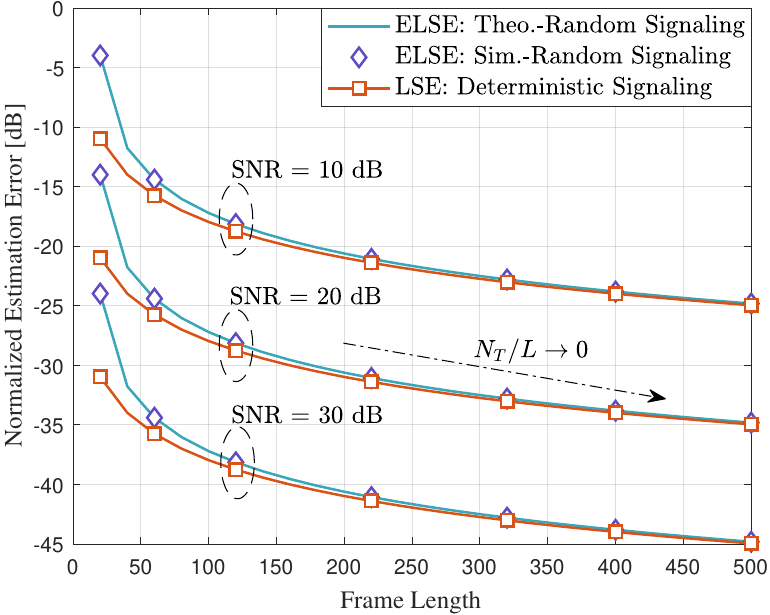} 
            }
\subfigure[S\&C performance tradeoffs.] { 
\label{ELSE_Tradeoff}
            \includegraphics[width=0.272\textwidth]{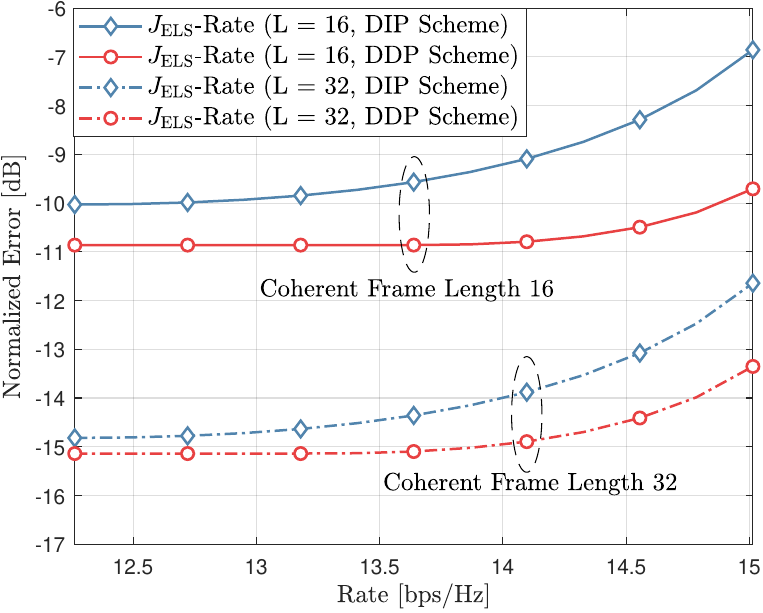} }
\caption{Spatial-domain MIMO precoding design in mono-static scenarios: System models and beneficial results.}
        \label{Precoding}
\end{figure*}

\subsubsection{Main Results} 
In Fig. \ref{64QAM_PCS}, we present the reshaped 64-QAM constellations under different kurtosis constraints. The standard 64-QAM constellation, characterized by a kurtosis of 1.38, exhibits both geometric and probabilistic symmetry. Considering the kurtosis value equals 1, the PCS cannot identify a constant modulus circle with unit power. Instead, it produces two constant modulus circles that are close to the unit modulus circle. In Fig. \ref{64QAM_Tradeoffs}, we illustrate the favorable performance tradeoffs of S\&C by the proposed algorithms. It is observed that the proposed PCS algorithms achieve flexible tradeoffs between S\&C, which is attained by adjusting the kurtosis of the constellations. Note that the 64-QAM PCS cannot attain the corner points under uniform PSK, since there is no potential constant-modulus shape under 64-QAM constellations. 

\subsubsection{Design Guidelines} 
By controlling the kurtosis of the constellation, the proposed PCS schemes may effectively approach the optimal tradeoff boundaries. To facilitate implementation, a look-up table of optimized constellation probabilities can be precomputed offline, allowing for rapid configuration based on specific ISAC performance requirements. This design enables efficient retrieval during transmission and offers flexibility to adapt to various ISAC scenarios, ensuring reliable performance in both S\&C.

\subsection{Spatial Domain: MIMO Precoding}

\subsubsection{Backgrounds} In ISAC scenarios, the precoding matrix needs to be optimized not only based on communication requirements but also for simultaneous sensing tasks under random signaling. Most existing works assume that the sample covariance matrix of random ISAC signals is asymptotically approximated to its statistical covariance matrix, which is deterministic. This assumption, however, is valid only when the frame length of the ISAC signal matrix greatly exceeds the array size. Such conditions may not hold in time-sensitive ISAC applications or large-scale MIMO-ISAC systems, leading to prohibitively high computational and signal processing costs for sensing tasks. Therefore, the ISAC transceivers need to explore short-frame observations to perform sensing tasks, where the inherent randomness of each signal frame realization may not be eliminated by the Law of Large Number \cite{LuTSP2024}.

\subsubsection{Models and Solutions} Consider the generic Gaussian linear model of mono-static ISAC systems, as illustrated in Fig. \ref{Precoding_Scenario}. The ISAC base station (BS) is assumed to sense the target impulse response (TIR) matrix, while simultaneously communicating with the multi-antenna user.
Considering whether the {\it a-priori} knowledge is available for the ISAC-BS, we define ergodic linear minimum mean square error (ELMMSE) and ergodic least squared error (ELSE) to quantify the average sensing performance under the celebrated least squared and LMMSE estimators, respectively. 
Notably, the sensing metrics under random ISAC signaling in both cases may be unified in the form of $\mathbb{E}_{\bm{S}}[f(\bm{W}; \bm{S})]$, which is aligned with the discussion in Sec. \ref{Sec2}. Here, $\bm{S}$ denotes the random ISAC signals, $\bm{W}$ denotes the to-be-optimized precoding matrix, and $f(\cdot)$ represents the instantaneous estimation error, i.e., LMMSE in Case 1 and LSE in Case 2. We first develop a pair of novel precoding schemes, namely, data-dependent precoding (DDP) and data-independent precoding (DIP), where the precoder changes adaptively based on the instantaneous random data streams, and remains unchanged for all realizations of $\bm{S}$, respectively. Notably, the DDP precoder attains favorable sensing performance, albeit at the cost of high complexity. The DIP may achieve a tradeoff between sensing performance and complexity \cite{LuTSP2024}. As a step further, both DIP and DDP schemes under sensing-optimal scenarios may be extended to ISAC scenarios, by explicitly constraining the communication rate \cite{LuTSP2024}.

\subsubsection{Main Results} We consider a multi-antenna system where the ISAC-BS is equipped with 16 transceiver antennas and the communication user (CU) is equipped with 8 receive antennas. The ISAC signal matrix follows a complex Gaussian distribution. As illustrated by Fig. \ref{ELSE_SensingOnly}, the ELSE attains the LSE with the increase of coherent frame length, highlighting the sensing performance degradation caused by random ISAC signaling. It is worth noting that the randomness may be asymptotically neglected with the increasing ratio of the frame length to the number of transmit antennas. In Fig. \ref{ELSE_Tradeoff}, both the proposed DDP and DIP schemes achieve favorable S\&C tradeoffs, albeit at the cost of increased complexity in employing the DDP method.

\subsubsection{Design Guidelines} 

The primary difference between the classical estimation metrics (e.g., LSE and LMMSE) and the average sensing performance metrics (e.g., ELSE and ELMMSE) lies in their applicability to scenarios involving random signals. For more general cases, the computation of $\mathbb{E}_{\bm{S}}[f(\bm{W}; \bm{S})]$ inherently accounts for the signal randomness and adheres to the statistical principle of expectation-based error evaluation.
We develop a general precoding framework under random signaling, where the proposed precoding schemes may be readily generalized to other ISAC scenarios, such as mobile perceptive networks, where the targets need to be sensed under short-frame signaling. On one hand, the innovative DDP scheme relies on perfect knowledge of each realization of random data symbols. Since the DDP matrix needs to adapt to the random data symbols in real time, achieving optimized ISAC performance at the expense of increased computational complexity. On the other hand, the DIP method may be implemented offline, offering a balanced tradeoff between ISAC performance and complexity.

\section{Open Problems and Future Directions}\label{Sec4}

Reusing random communication signals for sensing may achieve maximized resource efficiency without degrading communication performance. Nonetheless, many challenges still need to be addressed to comprehensively enhance sensing performance under the random signaling framework. In this section, we provide a brief discussion of future directions on random ISAC signal processing.

\subsection{2D AF Characterization}

We shed light on the ACF, namely, the zero-Doppler slice of the AF in Sec. \ref{Sec3}, which primarily addresses ranging performance. In 6G perceptive networks, diverse mobility scenarios require joint characterization of target range and velocity \cite{Zhiqing_netw,marwa_tuts}. To this end, future research should investigate the two-dimensional (2D) AF under arbitrary ISAC signaling to enable comprehensive delay-Doppler performance analysis. The 2D AF needs to be derived from the 2D ACF in the time-frequency domain, where the randomness of the communication data imposes significant challenges in the analytical characterization of the AF. Moreover, practical frame structure design, including the placement of fixed pilots and guard intervals, is crucial to ensuring reliable 2D AF performance.

\subsection{Channel-Coded ISAC Signaling}

Existing studies typically assume that random data are drawn from predefined i.i.d. constellations, overlooking the effects of channel coding. In essence, channel coding introduces redundancy into the original bit streams, which are correlated with the original random data. Therefore, when incorporating channel coding, e.g., convolutional coding, it is essential to analyze sensing performance under non-i.i.d. constellation symbols. Future research should also investigate the potential benefits of existing channel coding schemes (e.g., LDPC and Polar codes) for improving both S\&C performance.

\subsection{Distributed ISAC Networks}

Distributed ISAC networks leverage multiple transceiver nodes to achieve improved S\&C performance. However, those non-cooperative nodes may lack knowledge about the random ISAC symbols at the sensing Rx. This absence of symbol information may result in mismatched filtering errors, degrading the average sensing performance. To address this issue, the so-called ``decoding-then-sensing'' framework may be further developed, where the sensing receiver first estimates the received communication symbols and then uses the estimated results to perform matched filtering with the whole echo signals. Future directions should analytically derive the relationship between symbol decoding errors and the matched filter output. It is promising to utilize such a framework to facilitate the investigation of optimal resource allocations of distributed ISAC networks.

\section{Conclusion}

In this article, we present a general signal processing framework for sensing under random ISAC signaling. We commence with formulating the sensing performance metrics and the optimal ranging waveform when using random communication signals. Building on this, we systematically explore three key technologies to improve the average sensing performance: time-domain pulse shaping, frequency-domain constellation shaping, and spatial-domain precoding techniques. Finally, we conclude by outlining future research directions in random ISAC signal processing.

\bibliographystyle{IEEEtran}
\bibliography{ref}

\end{document}